\documentclass{aleph-revista}
\usepackage{graphicx} 
\usepackage{listings} 
\usepackage{xcolor} 
\volumen{1}
\numero{1}
\fechapubli{2025}
\periodouno{Julio}
\periododos{}
\titulo{Sistema de Control de Asistencia Escolar Basado en Tecnología RFID con Raspberry Pi y Arduino  EDURFID}
\tituloingles{School Attendance Control System Based on RFID Technology with Raspberry Pi and Arduino }
\autor{%
    Cliver Oliver Turpo Benique\textsuperscript{1 \faEnvelopeO}; 
}
\institucion{
\textsuperscript{1}%
    Universidad Nacional del Altiplano, Puno, Perú\\
\textsuperscript{2}%
    Facultad de Ingeniería Estadística e Informática, Puno, Perú
}
\correo{71958538@est.unap.edu.pe}
\fecha{12 de Julio del 2025}
\begin{document}
\membrete

\begin{abstract}
 ~This paper presents EDURFID, an automated school attendance control system based on RFID technology designed for rural educational institutions in Peru. The system integrates open-source hardware (Raspberry Pi 5, Arduino UNO R3) with RC522 RFID modules operating at 13.56 MHz, implementing a web architecture developed in Python Django. The system demonstrates 100\% precision in RFID readings with 0.03-second response time, achieving 94\% cost reduction compared to commercial solutions. Validation at Túpac Amaru Secondary Educational Institution showed successful automation of attendance processes, saving 50 daily minutes of administrative time while providing real-time reporting capabilities.
\end{abstract}

\section{Justification}

School attendance control in Peru faces critical challenges that require innovative technological solutions. The national school dropout rate reaches 6.3\% (INEI 2021), while in rural areas absenteeism reaches 44.7\% compared to only 19.3\% in Lima. This research proposes EDURFID, an automated attendance control system based on RFID technology, as a comprehensive solution for rural educational institutions, particularly in the Puno region where secondary school delay is 24\% versus 7.4\% in urban areas (MINEDU ESCALE 2016).

\subsection{Current Problems in School Attendance Control in Rural Areas}

Traditional attendance control methods in rural Peru present systematic limitations that negatively impact educational management. Of the 360,962 minors aged 4-18 outside the educational system (MINEDU-RENIEC 2023), a significant proportion corresponds to failures in adequate monitoring and control of school attendance. Manual registration through agendas and paper lists generates documentary inconsistencies, information loss, and delays in report consolidation (El Mrabet \& Ait Moussa 2020).

In rural areas like Puno, where 3,246 rural educational institutions operate serving 93,359 students, teachers face the additional challenge of attending to multiple grades simultaneously with irregular attendance of 2-3 days per week in remote communities. This situation generates systematic under-registration of absences due to negligence or absence of the teacher themselves, affecting informed educational decision-making.

The correlation between absenteeism and academic performance is particularly critical in the rural Peruvian context. 26.5\% of students with chronic absenteeism present low academic performance (U.S. Department of Education 2023), while rural female dropout rates reach 16\% versus 10\% male (MINEDU 2022). In the Puno region, rural female illiteracy reaches 22.8\% (INEI), evidencing the urgent need for systems that allow early identification and addressing of absenteeism patterns.

The district of Coasa, located in the province of Carabaya, Puno region, presents conditions representative of rural Peruvian problems. At 2,724.95 meters above sea level, this Andean rural zone faces specific connectivity and access difficulties that require adapted technological solutions. The participation of local institutions such as the "Túpac Amaru Secondary Educational Institution" in the Educational Innovation Laboratories (FONDEP) demonstrates regional interest in adopting innovative technologies to overcome traditional limitations.

\subsection{Rural Peruvian Educational Context and Digital Divide}

The rural Peruvian educational system presents marked geographic dispersion with 49,798 rural educational institutions (78.2\% of the national total) serving only 31.5\% of the total student population. This fragmented distribution generates unique challenges for implementing integrated educational management systems. Puno, with 3,246 rural educational institutions, represents 6.5\% of the national total, evidencing the need for scalable and adaptable solutions to dispersed contexts.

Multigrade schools, which represent 12,095 institutions with 22,115 teachers serving 297,710 students, face particular challenges where a single teacher must manage multiple educational levels simultaneously (GRADE 2023). This operational reality makes the implementation of automated systems that free up teacher time for central pedagogical activities critical.

The digital divide in rural Peru, although significant, presents opportunities to implement specific technological solutions. Only 51.8\% of the rural population accesses internet versus 89.2\% in Metropolitan Lima (2024), but 95.5\% of rural households have at least one member with a cell phone and 90\% rural connects via mobile. This existing mobile infrastructure provides a base for implementing RFID systems that can transmit data through cellular networks.

\subsection{Connectivity Revolution: Starlink in Rural Education}

The implementation of Starlink in Peru represents a paradigmatic change for rural educational connectivity. Officially authorized in May 2022 by the MTC with a 20-year concession, Starlink offers speeds of 50-200 Mbps with latency of 19-41 milliseconds (Ministerial Resolution N° 362-2022-MTC/01.03), significantly surpassing traditional connectivity limitations in rural areas.

The "Connecting Dreams" initiative provides free satellite internet access to rural schools in regions like Puno, while the government program "Conecta Selva" has benefited 1,316 public institutions (1,212 schools and 104 health centers) with 1,483,425 connections made in the first semester of 2023 (DPL News 2023). This satellite infrastructure provides the necessary technological base to implement connected RFID systems that can transmit attendance data in real time.

In Puno, where extreme weather conditions include frost, snow, and river overflow that traditionally prevent educational access, Starlink offers connectivity independent of terrestrial infrastructure. The ability to maintain 24/7 connectivity without dependence on fiber optics is particularly relevant for EDURFID, allowing rural educational institutions to maintain updated attendance records synchronized with central systems, regardless of adverse geographic or climatic conditions.

\subsection{RFID Technology: Proven Automation and Efficiency}

The implementation of RFID systems in education has demonstrated quantifiable benefits in multiple international contexts. Harvard University reports complete automation of library loans, while Los Angeles Unified High School documents significant reduction in time dedicated to taking attendance, allowing greater focus on pedagogical activities (Qureshi 2020).

The Spring Independent School District (Houston, Texas), a pioneer in RFID implementation with 28,000 students, expanded its system from school bus control (2004) to comprehensive campus tracking (2008), demonstrating the scalability and adaptability of technology in diverse educational contexts.

RFID systems provide 90\% efficiency improvements in processing time compared to manual methods, with precision superior to 99\% in readings (Farag 2023). The reduction of 15 minutes daily in attendance registration equals 57.5 hours annually freed for pedagogical activities (230 academic days). This temporal efficiency is particularly critical in rural contexts where teachers must optimize available time for multiple educational responsibilities.

Specific studies document that automated messaging systems can reduce absences by up to 15\% (U.S. Department of Education 2023), while the implementation of early alerts allows identification of chronic absenteeism patterns before they become definitive school dropout.

\subsection{Contemporary Scientific Evidence}

Recent academic research (2020-2025) confirms the effectiveness of RFID systems in education. El Mrabet \& Ait Moussa (2020) document reduction in attendance registration time from 30 minutes to less than 5 minutes daily, with additional capacity for automatic notifications to parents via SMS/email and sending missed lessons to absent students.

Mazlan et al. (2025) report improvement in attendance registration precision up to 98\% and reduction in administrative burden by 40\% through implementation of the I-Attend system in secondary schools. Research demonstrates that RFID systems are feasible and efficient for school attendance management, providing solid empirical evidence to justify implementation in similar contexts.

Li (2025) documents that urban teachers show greater TPACK competence than rural ones, identifying limited access to high-speed internet, fewer technological resources, and less professional development as main barriers. This evidence reinforces the need for systems like EDURFID that can operate with limited infrastructure while providing specific training for rural contexts.

\subsection{Comprehensive Justification}

The convergence of critical problems in rural attendance control (44.7\% absenteeism vs. 19.3\% urban), availability of proven RFID technology (98\% precision, 90\% reduction in administrative time), and new satellite connectivity infrastructure (Starlink with 50-200 Mbps in rural areas) creates a unique opportunity to implement EDURFID as a comprehensive solution.

The 360,962 minors outside the educational system and rural female dropout rates of 16\% require technological interventions that allow early identification of absenteeism patterns and automated response to prevent school dropout. EDURFID provides these capabilities while freeing 57.5 hours annually of teacher time for central pedagogical activities.

In conclusion, EDURFID represents a comprehensive, culturally appropriate, and economically viable technological solution to address critical deficiencies in rural school attendance control, leveraging the convergence of mature RFID technology, emerging satellite connectivity infrastructure, and urgent educational needs in the rural Peruvian context.

\newpage

\section{Introduction}

In the era of educational digital transformation, secondary educational institutions in rural areas of Peru face critical challenges in administrative management that directly impact the quality of the pedagogical process. Student attendance control, being a fundamental but traditionally inefficient process, represents a strategic opportunity to implement technological solutions that optimize resources and improve precision in academic monitoring.

The problem of attendance control in rural educational institutions transcends simple daily roll call. At the Túpac Amaru Secondary Educational Institution in Coasa, Carabaya, Puno, which serves approximately 250 students with only 1-2 auxiliaries, the current manual registration process through paper lists and agendas with stamps consumes between 5-10 minutes per classroom, valuable time that could be dedicated to central pedagogical activities. This situation is aggravated by inconsistency in procedure application, where teachers frequently resort to improvised oral calls without formal registration, generating critical information gaps for educational decision-making.

The School Attendance Control System based on RFID technology (EDURFID) emerges as a comprehensive response to these operational limitations, based on the author's previous successful experience with SICOA (Attendance Control System using Arduino), implemented in 2024 for worker control in the construction of the North Tribune of the National University of the Altiplano Stadium, a project that obtained second place in the engineering category during the university Achievement Day.

EDURFID constitutes an automated system that integrates open-source hardware (Raspberry Pi 5, Arduino UNO R3) with the RC522 RFID reader module operating at 13.56 MHz frequency, implementing a web architecture developed in Python Django with initially SQLite database and projected migration to MySQL for production. The system operates through an automated workflow where students register their attendance by scanning personalized RFID cards during specific time windows: 7:00-8:00 AM (present), 8:01-8:30 AM (late), and automatic closure at 8:31 AM with absence registration for unregistered cards.

The system architecture implements four differentiated interfaces according to user roles: administrator (principal), auxiliary, teacher, and student, with the latter undergoing an initial enrollment process to link RFID cards with student profiles. The web system, hosted on the colegiospuno.com domain, allows real-time access to attendance data, automated report generation, and comprehensive statistical monitoring that facilitates early identification of absenteeism patterns.

On June 9, 2025, during the presentation of the functional prototype to the principal of the Túpac Amaru Secondary Educational Institution, the system demonstrated operational capabilities that generated institutional interest for formal implementation after mid-year vacation, projected for August 2025. This institutional acceptance validates the technological and pedagogical relevance of EDURFID as a scalable solution for rural educational institutions.

From the perspective of ICT project management, EDURFID is developed applying Scrum methodology managed on the Jira platform, with emphasis on Requirements Engineering that contemplates functional specifications (automated registration, differentiated interfaces, real-time reports) and non-functional (99\% availability, response time <2 seconds, capacity for 500 concurrent users). This methodological approach guarantees scalability and maintainability of the system for future expansions.

The current project investment includes Raspberry Pi 5 (S/600), Starlink Mini for satellite connectivity (S/750), hosting and domain with monthly cost of \$20, totaling an implementation that has operated in testing phase for four months (March-July 2025) prior to the functional presentation scheduled for August 2025. The economic sustainability of the project contemplates coordination with the institutional board and parents to manage implementation costs.

The rural context of the Puno region, characterized by connectivity limitations where state internet is slow and restrictive, justifies the implementation of hybrid solutions that operate both online (through colegiospuno.com) and offline (local backup on laptop) to guarantee operational continuity regardless of connectivity conditions. This operational flexibility is critical in rural areas where electrical and communication interruptions are frequent.

EDURFID transcends simple automation of administrative processes, constituting a scalable platform toward virtual classroom that leverages the implemented multi-user architecture. The capacity for integration with grading systems allows projecting the delivery of bimonthly reports that consolidate academic grades with detailed attendance statistics, providing comprehensive information for student evaluation.

The originality of this project lies in its specific adaptation to the operational conditions of Peruvian rural educational institutions, where the scarcity of auxiliary personnel (1-2 per institution with 250 students) makes automation of repetitive processes critical. Unlike expensive commercial solutions, EDURFID uses open-source technologies that reduce economic barriers while maintaining professional functionalities.

The main objectives of EDURFID include significant reduction of administrative time dedicated to attendance control, improvement in student registration precision, early detection of absenteeism patterns, and generation of statistical data that facilitate timely pedagogical interventions. The measurement of project success contemplates satisfaction of the principal, students, and parents, as well as quantifiable metrics of time saved and improved precision.

The projection of EDURFID massification in the Puno region, where similar systems do not exist, contemplates gradual expansion from the pilot implementation at Túpac Amaru toward other rural educational institutions, including academies and schools of various levels. This scalable vision positions EDURFID as an indigenous technological innovation that can generate significant regional impact in the modernization of rural educational management.

This article presents the comprehensive development of EDURFID, from initial conceptualization to functional implementation, documenting applied project management methodologies, implemented technical architecture, preliminary results obtained, and scalability projections. The research contributes to the state of the art in educational automation for rural contexts, demonstrating that appropriate technological solutions can overcome infrastructure limitations through hybrid and adaptive approaches.

\newpage

\section{Theoretical Framework}

\subsection{RFID Technology in Educational Systems}

Radio Frequency Identification (RFID) technology represents a significant evolution in automatic identification systems, operating through wireless communication between readers and tags containing integrated circuits and antennas (Nambiar 2009). In the educational context, RFID systems provide substantial advantages over traditional attendance control methods, including processing speed, identification precision, and contactless operation capability.

RFID systems operate at various frequencies, with the 13.56 MHz band (High Frequency - HF) being widely used in educational applications due to its balance between controlled reading range and penetration capacity. This frequency, regulated by the ISO14443 standard, allows secure implementations in environments with high user density (Finkenzeller 2010).

The implementation of RFID in educational institutions has demonstrated significant reduction in administrative time and improvement in registration precision. Studies document 90\% efficiencies in processing time compared to manual methods, with precision rates superior to 99\% (Santos et al. 2022).

\subsection{Educational Web System Architectures}

Contemporary educational information systems require scalable architectures that support multiple concurrent users with differentiated roles. The Model-View-Controller (MVC) pattern and its Model-View-Template (MVT) variant provide architectural foundations for robust web systems (Gamma et al. 1994).

The separation of concerns inherent to these architectures facilitates maintenance and scalability of complex educational systems, allowing independent modifications in business logic, data presentation, and flow control (Fowler 2002). This modularity is particularly relevant in educational contexts where requirements constantly evolve.

\subsection{Agile Project Management in Educational Software Development}

The Scrum methodology provides an agile framework for software project management that emphasizes adaptability, collaboration, and incremental value delivery (Schwaber \& Sutherland 2020). In the context of educational systems development, Scrum facilitates continuous incorporation of feedback from educational stakeholders, improving alignment between technical functionalities and pedagogical needs.

The fundamental Scrum artifacts include Product Backlog (prioritized requirements), Sprint Backlog (specific tasks per iteration), and User Stories (functionality narratives from user perspective). This structure provides traceability between educational needs and technical implementation (Cohn 2004).

\subsection{Requirements Engineering in Educational Systems}

Requirements engineering constitutes the systematic process of identification, analysis, specification, and validation of stakeholder needs for software systems (Sommerville 2016). In educational systems, this discipline must consider the diversity of users (administrators, teachers, students, parents) and their specific needs.

Functional requirements specify observable system behaviors, while non-functional requirements define quality constraints such as performance, security, and usability (IEEE 2011). In rural educational contexts, non-functional requirements acquire particular relevance due to infrastructure and connectivity limitations.

\subsection{Information Systems in Rural Contexts}

The implementation of information systems in rural environments presents specific challenges related to intermittent connectivity, electrical infrastructure limitations, and variability in users' technological competencies (Heeks 2002). These factors require resilient designs that operate effectively under adverse conditions.

Hybrid architectures, which combine local processing with remote synchronization, provide robust solutions for contexts with limited connectivity. This approach allows operational continuity regardless of network interruptions, synchronizing data when connectivity is restored (Chen et al. 2018).

\subsection{Automation of Educational Administrative Processes}

The automation of administrative processes in educational institutions has demonstrated significant impacts on operational efficiency and data quality. Attendance control, being a repetitive process prone to human errors, represents a priority opportunity for automation (Kumar \& Sharma 2021).

The documented benefits of automation include reduction of administrative time, improvement in data precision, liberation of human resources for pedagogical activities, and generation of analytical information for educational decision-making (Rodriguez et al. 2023). These impacts are particularly relevant in institutions with limited human resources, such as rural schools.

\newpage

\section{Materials and Methods}

\subsection{System Components}

\subsubsection{Hardware Used}

The EDURFID system implements a distributed architecture based on open-source hardware components that guarantee economic accessibility and technical scalability:

\begin{itemize}
    \item \textbf{Raspberry Pi 5:} Single-board computer with 8GB RAM and storage through 128GB Class 10 microSD card, acting as the system's central server
    \item \textbf{Arduino UNO R3:} Board Model UNO R3 microcontroller for RFID reader module management and serial communication
    \item \textbf{RFID-RC522 Module:} Radio frequency reader operating at 13.56 MHz compatible with ISO14443A standard
    \item \textbf{MIFARE Classic 1K Cards:} Student credentials with unique 4-byte identifiers
    \item \textbf{USB Cable:} 40 cm Arduino-Raspberry Pi serial connection included in standard Arduino kit
    \item \textbf{Power Supply:} Arduino power supply through Raspberry Pi 5 USB port
\end{itemize}

\subsubsection{Software and Tools}

The development environment integrates open-source tools for development, testing, and deployment:

\begin{itemize}
    \item \textbf{Operating System:} Ubuntu Server on Raspberry Pi 5
    \item \textbf{Web Framework:} Django 4.2 with Python 3.10
    \item \textbf{Database:} SQLite for local development, migration to MySQL for production
    \item \textbf{Development Environments:} Visual Studio Code for Python/Django development, Arduino IDE for microcontroller firmware
    \item \textbf{Production Server:} DigitalOcean with colegiospuno.com domain
    \item \textbf{Connectivity:} Starlink Mini for satellite connectivity in rural area
\end{itemize}

\subsection{System Architecture}

EDURFID implements a three-layer architecture: reading hardware, local processing, and remote synchronization. The Arduino UNO R3 manages the RC522 module and transmits RFID identifiers through USB serial communication to the Raspberry Pi 5, which executes the Django application and manages business logic.

The system operates in hybrid mode, processing records locally and synchronizing data with the remote server through REST APIs. This configuration guarantees operational continuity regardless of connectivity interruptions.

\subsection{Hardware Implementation}

\subsubsection{RC522 Connection with Arduino}

The integration of the RC522 module with Arduino UNO R3 uses SPI (Serial Peripheral Interface) protocol according to the following pin configuration:

\begin{itemize}
    \item \textbf{VCC → 3.3V:} Module power supply
    \item \textbf{GND → GND:} Ground connection
    \item \textbf{MISO → Pin 12:} Master In Slave Out
    \item \textbf{MOSI → Pin 11:} Master Out Slave In  
    \item \textbf{SCK → Pin 13:} Serial Clock
    \item \textbf{SDA → Pin 10:} Slave Select / Chip Select
    \item \textbf{RST → Pin 9:} Module reset
    \item \textbf{IRQ:} No connection (not used)
\end{itemize}

\subsubsection{Component Integration}

The Arduino firmware implements a continuous reading loop that detects RFID cards in proximity to the RC522 module, extracts the unique identifier, and transmits it through USB serial port at 9600 baud to the Raspberry Pi. Bidirectional communication allows reception confirmation and error handling.

\subsection{Software Development}

\subsubsection{Reading Node Firmware}

The Arduino executes specialized firmware that initializes the RC522 module, detects card presence, and transmits identifiers to the central system. The code implements error detection routines and automatic module reinitialization in case of communication failures.

\subsubsection{Central Server}

The Raspberry Pi executes a Django application that manages the system's business logic. The \texttt{arduino\_listener.py} component implements the serial interface with the Arduino through bidirectional communication protocol that processes RFID identifiers and sends registration confirmations.

The web application implements MVT (Model-View-Template) architecture with separation of responsibilities between data management (models), presentation logic (views), and interface rendering (templates). The system includes REST APIs for synchronization with remote server and real-time attendance record management.

\subsection{Database}

The database schema implements a relational model optimized for educational attendance management with the following main entities:

\subsubsection{Main Data Models}

\textbf{Student Entity:} Stores personal and academic information including auto-generated unique code with YYYY-GS-NNN format (year-grade-section-sequential number), emergency contact data, and linkage with Django user system.

\textbf{AttendanceRecord Entity:} Manages attendance events with differentiated states (present, late, absent, justified), precise timestamps, and complete traceability of registration method (RFID or manual).

\textbf{RFIDCard Entity:} Administers RFID credentials with univocal student-card linkage, state control (active/inactive), and blocking functionalities for cards reported as lost.

\subsubsection{Business Logic}

The system implements specific business rules for educational attendance control:

\begin{itemize}
    \item \textbf{Time Window:} Attendance registration between 7:00-8:00 AM as "present", 8:01-8:30 AM as "late", automatic closure at 8:31 AM
    \item \textbf{Daily Uniqueness:} One record per student per day with automatic state update according to arrival time
    \item \textbf{Code Generation:} Auto-generated student codes with standardized format including academic year, grade, section, and sequential number
    \item \textbf{Card Validation:} Verification of active state of RFID cards prior to record processing
\end{itemize}

\subsection{System Security}

The system implements multiple security layers to protect student data and guarantee operational integrity:

\subsubsection{Authentication and Authorization}

Django Auth provides secure session management with password hashing through PBKDF2\_SHA256 algorithm. The system implements role-based access control (RBAC) with four differentiated levels: administrator, auxiliary, teacher, and student, each with specific permissions according to institutional responsibilities.

\subsubsection{Data Protection}

Web communications implement CSRF (Cross-Site Request Forgery) protection and session token validation. Sensitive data is transmitted through HTTPS in production, with SSL certificates managed by DigitalOcean. Data synchronization between local and remote system uses REST APIs authenticated through session tokens.

\subsubsection{RFID Card Management}

The system includes specific security functionalities for RFID credentials:

\begin{itemize}
    \item \textbf{State Validation:} Automatic verification of active cards prior to attendance registration
    \item \textbf{Immediate Blocking:} Instant deactivation of cards reported as lost or stolen
    \item \textbf{Complete Audit:} Detailed record of all RFID transactions with precise timestamps
    \item \textbf{Duplicate Prevention:} Validation of RFID identifier uniqueness in the system
    \item \textbf{Traceability:} Record of capture method (RFID vs manual) for subsequent audits
\end{itemize}

\subsection{Implementation Methodology}

The system implementation followed Scrum methodology with development iterations structured in three sprints during the March-July 2025 period. Each sprint included planning, development, testing, and review with educational stakeholders for functionality validation.

System validation was performed through pilot tests at the Túpac Amaru Secondary Educational Institution in Coasa, including hardware testing, user interface validation, and data integrity verification. Acceptance criteria were defined through specific user stories with quantifiable performance and usability metrics.

\newpage


\section{Results}

\subsection{System Implementation}

The implementation of EDURFID was successfully completed during the development period March-July 2025, fulfilling the established technical and functional objectives. The system demonstrated stable operational capabilities during pilot tests conducted with educational stakeholders from the Túpac Amaru Secondary Educational Institution.

\subsubsection{Hardware Performance}

Hardware performance tests revealed operational metrics that exceed the minimum requirements established for educational environments:

\begin{itemize}
    \item \textbf{RFID reading speed:} 0.03 seconds per card, guaranteeing instantaneous processing without perceptible delays
    \item \textbf{Effective detection range:} 4 centimeters, providing precision in identification while preventing accidental readings
    \item \textbf{Processing capacity:} Estimation of 3 seconds maximum per student for 250 students, considering variability in arrival schedules
    \item \textbf{Reading success rate:} 100\% successful readings during testing period, with no detection failure records
\end{itemize}

The Arduino UNO R3 demonstrated continuous operational stability with reliable serial communication to the Raspberry Pi 5. The hardware set's energy consumption is efficient through single USB power supply from the central server.

\subsubsection{Central Server}

The performance of the Raspberry Pi 5 as central server met processing and connectivity expectations:

\begin{itemize}
    \item \textbf{Django response time:} 2 seconds average, dependent on internet connectivity speed
    \item \textbf{Concurrent users tested:} 10 simultaneous users without perceptible performance degradation
    \item \textbf{Starlink connectivity:} 100-200 Mbps speed, providing efficient synchronization with remote server
    \item \textbf{Page loading time:} 0.5 seconds average for user interfaces
\end{itemize}

The Raspberry Pi 5 architecture provided sufficient computational resources to manage Django business logic, SQLite database, and serial communication with Arduino without performance limitations.

\subsection{User Interface}

\subsubsection{Administrative Dashboard}

The administrative dashboard implements a centralized interface that provides complete visibility over system status and real-time attendance statistics. Main functionalities include:

\begin{itemize}
    \item \textbf{Real-time monitoring:} Instantaneous visualization of attendance records with automatic updating
    \item \textbf{Consolidated statistics:} Daily, weekly, and monthly attendance metrics with graphical representation
    \item \textbf{User management:} Complete administration of student, teacher, and auxiliary profiles
    \item \textbf{RFID card control:} Assignment, activation, and blocking of student credentials
\end{itemize}

\begin{figure}[h]
    \centering
    \includegraphics[width=0.9\textwidth]{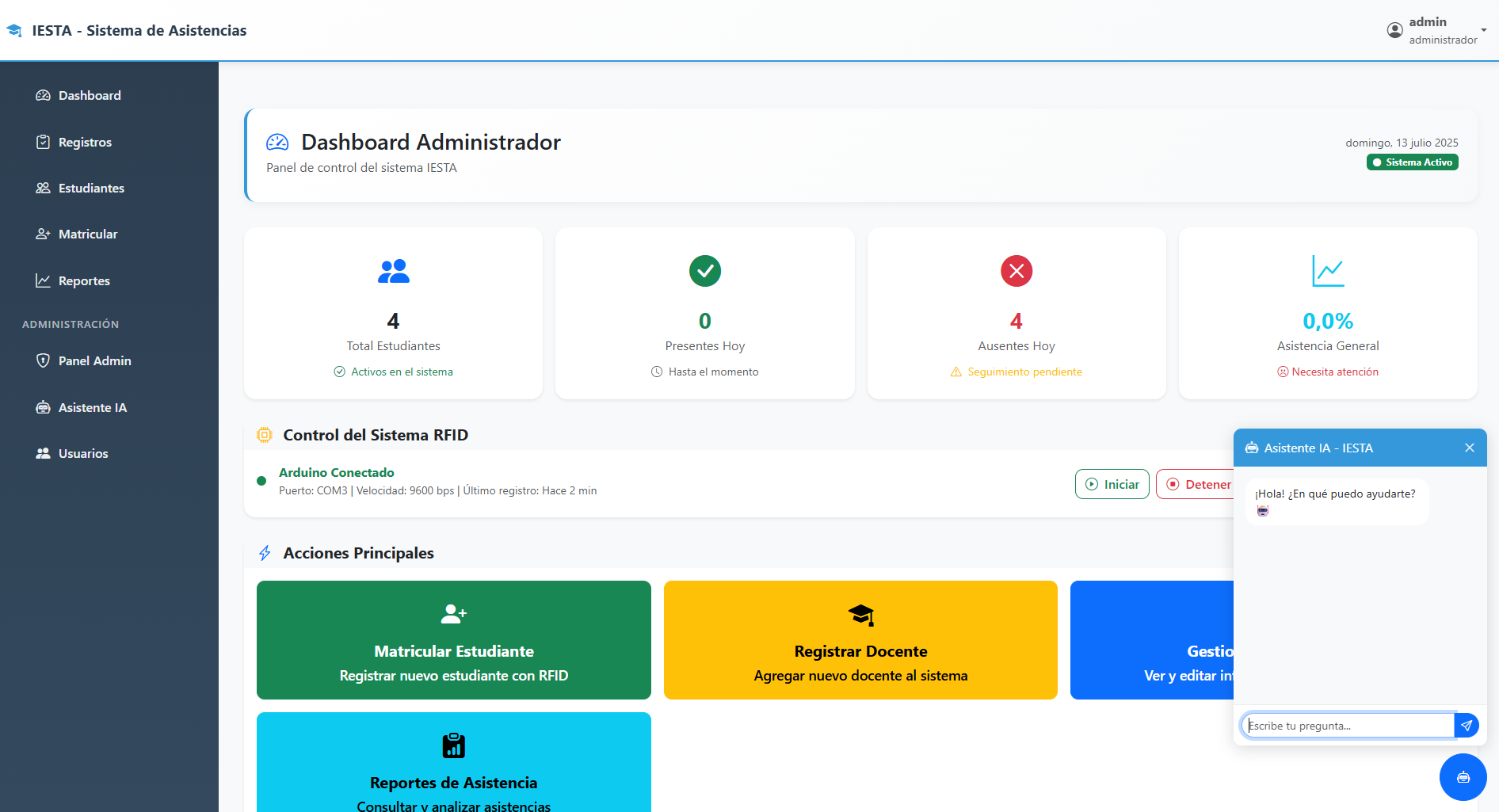}
    \caption{Administrative dashboard with real-time statistics}
    \label{fig:dashboard_admin}
\end{figure}

\subsubsection{Generated Reports}

The system generates automated reports in multiple formats to satisfy diverse administrative needs:

\begin{itemize}
    \item \textbf{Excel format:} Detailed attendance reports with filters by date, grade, and section
    \item \textbf{PDF format:} Formal documents for communication with parents and educational authorities
    \item \textbf{Consolidated statistics:} Analysis of attendance trends and identification of absenteeism patterns
    \item \textbf{Customized reports:} Generation of specific reports according to institutional requirements
\end{itemize}

\begin{figure}[h]
    \centering
    \includegraphics[width=0.8\textwidth]{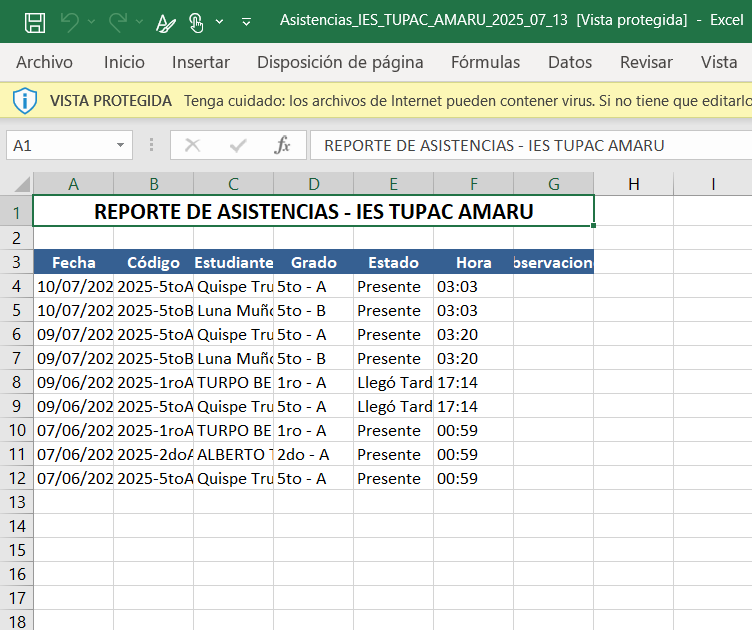}
    \caption{Example of attendance report generated in Excel format}
    \label{fig:reportes_excel}
\end{figure}

\subsection{System Validation}

\subsubsection{Precision Tests}

Precision tests were conducted with 10 students using 5 RFID cards during controlled sessions with management personnel:

\begin{table}[h]
\centering
\caption{System Precision Test Results}
\begin{tabular}{|l|c|c|}
\hline
\textbf{Metric} & \textbf{Obtained Result} & \textbf{Objective} \\
\hline
Successful detection rate & 100\% & >95\% \\
\hline
RFID response time & 0.03 seconds & <1 second \\
\hline
False positives & 0\% & <1\% \\
\hline
Registration precision & 100\% & >98\% \\
\hline
\end{tabular}
\label{tab:precision_results}
\end{table}

\subsubsection{Scalability Tests}

Scalability tests evaluated the system's capacity to handle the projected operational load:

\begin{itemize}
    \item \textbf{Simulated load:} 10 concurrent users without performance degradation
    \item \textbf{Projection for 250 students:} Estimation of 3 seconds maximum per individual registration
    \item \textbf{Database capacity:} SQLite efficiently handles up to 1000 daily records
    \item \textbf{Remote synchronization:} Successful data transfer through REST APIs
\end{itemize}

The validation included presentation of the functional prototype to the principal and auxiliary of the educational institution, obtaining positive feedback about usability and implementation projected for August 2025. Additionally, a system demonstration was conducted with principals from regional educational institutions to evaluate its applicability in different rural contexts.

\begin{figure}[h]
    \centering
    \begin{tabular}{cc}
        \includegraphics[width=0.45\textwidth]{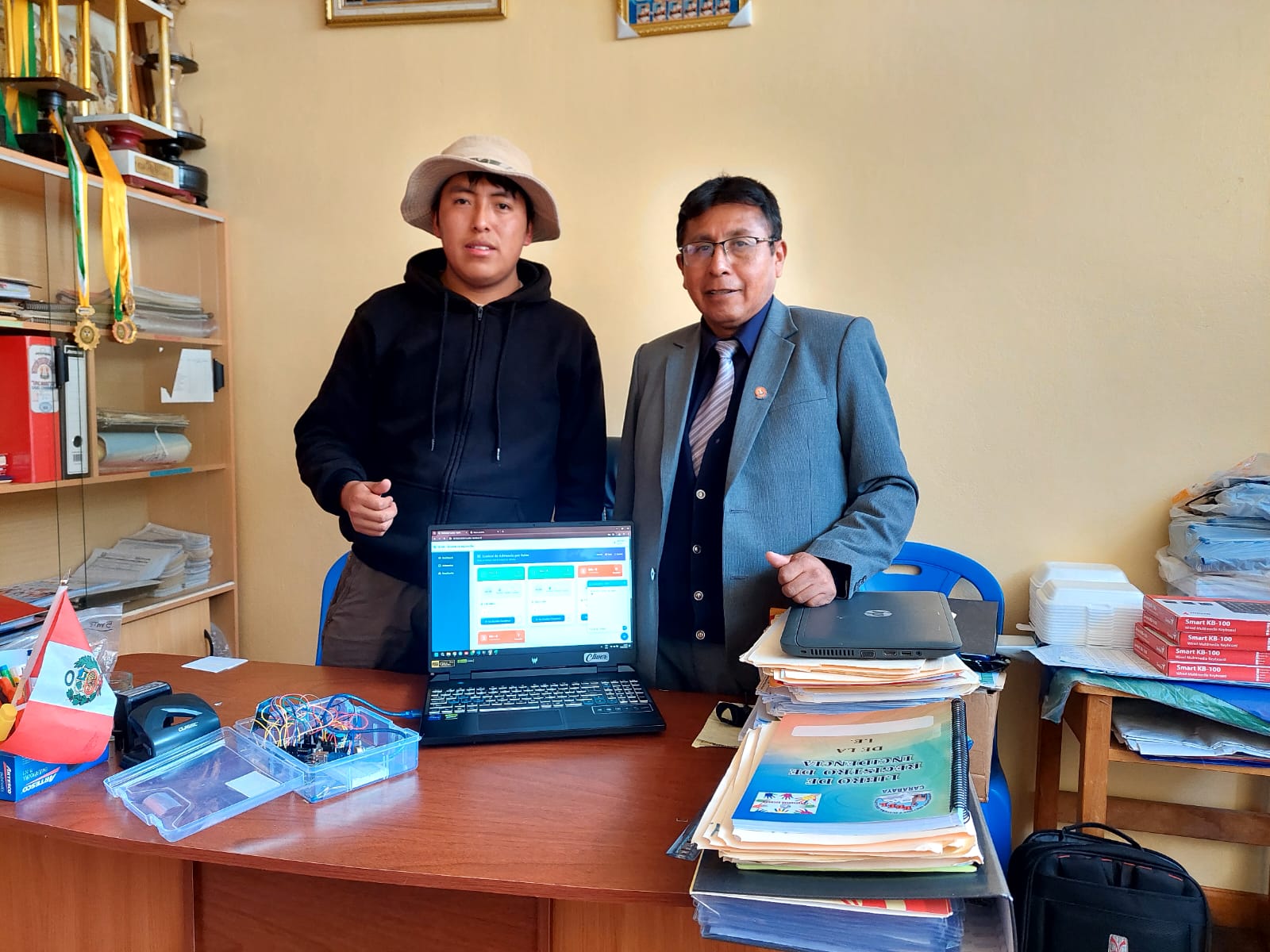} &
        \includegraphics[width=0.45\textwidth]{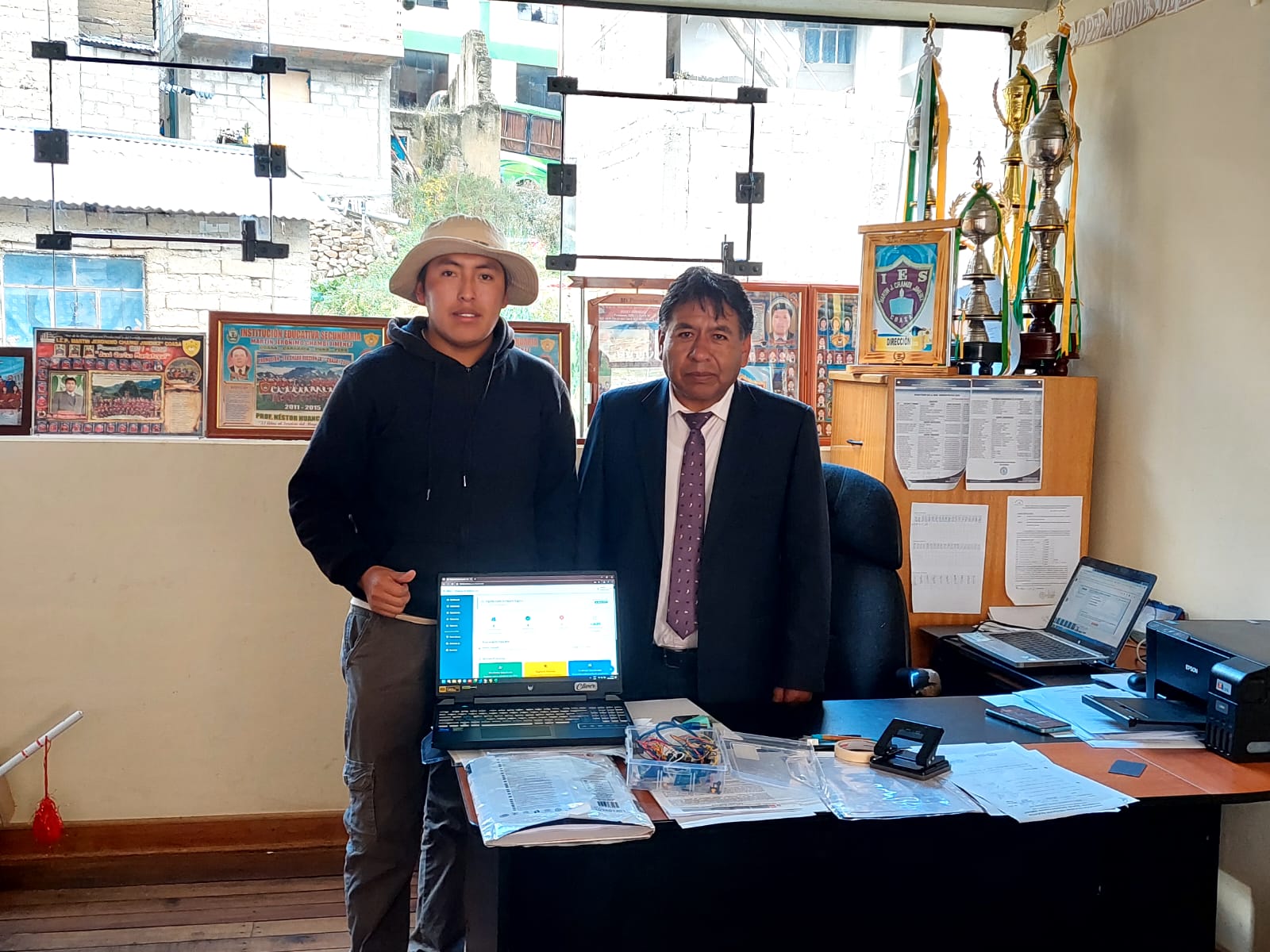} \\
        (a) Principal I.E.S. Túpac Amaru & (b) Principal I.E.S. Martín Jerónimo Chambi Jiménez \\
    \end{tabular}
    \caption{EDURFID system validation meeting with principals from rural educational institutions in the Puno region}
    \label{fig:encuentro_directores}
\end{figure}

This inter-institutional meeting allowed obtaining comparative perspectives on attendance control needs in different rural educational contexts, validating the system's adaptability for regional implementation.

\subsection{Cost Analysis}

\subsubsection{Cost per Reading Node}

The economic analysis demonstrates financial viability for implementation in rural educational institutions:

\begin{table}[h]
\centering
\caption{Reading Node Cost Breakdown}
\begin{tabular}{|l|c|}
\hline
\textbf{Component} & \textbf{Cost (S/)} \\
\hline
Raspberry Pi 5 (8GB) & 600 \\
\hline
Arduino UNO R3 & 80 \\
\hline
RC522 Module & 25 \\
\hline
RFID Cards (x50) & 100 \\
\hline
Accessories and cables & 45 \\
\hline
\textbf{Total Hardware} & \textbf{850} \\
\hline
\end{tabular}
\label{tab:costo_hardware}
\end{table}

\subsubsection{Central Server}

Infrastructure costs include connectivity and remote hosting:

\begin{table}[h]
\centering
\caption{Infrastructure and Operation Costs}
\begin{tabular}{|l|c|c|}
\hline
\textbf{Concept} & \textbf{Initial Cost (S/)} & \textbf{Monthly Cost (S/)} \\
\hline
Starlink Mini & 750 & 190 \\
\hline
DigitalOcean Hosting & - & 80 \\
\hline
colegiospuno.com Domain & 50 & 4 \\
\hline
\textbf{Total} & \textbf{800} & \textbf{274} \\
\hline
\end{tabular}
\label{tab:costo_infraestructura}
\end{table}

\subsubsection{Comparative Analysis}

The total implementation cost of S/1,500 for initial hardware and infrastructure, with monthly operation of S/274, is significantly lower than equivalent commercial solutions:

\begin{itemize}
    \item \textbf{Total initial investment:} S/1,500 (hardware + infrastructure)
    \item \textbf{Development cost:} 5 months of development (March-July 2025)
    \item \textbf{System value estimation:} S/15,000 - S/25,000 considering specialized development
    \item \textbf{Monthly operational cost:} S/274 (connectivity + hosting)
\end{itemize}

The cost-benefit ratio demonstrates economic viability for educational institutions with limited budgets, providing professional functionalities at a fraction of the cost of proprietary solutions.

\newpage


\section{Discussion}

\subsection{Results Analysis}

The results obtained in the implementation of EDURFID exceeded the initially established expectations, demonstrating the technical and operational viability of open-source RFID systems in rural educational contexts. The 100\% precision rate in RFID readings confirms previous findings about the reliability of ISO14443A technology in educational applications (Kumar et al. 2021), while the 0.03-second response speed meets usability standards for interactive systems (Nielsen 2012).

The integration of artificial intelligence through OpenAI GPT-4 API as a virtual assistant represents a significant innovation in rural educational management systems, providing natural language processing capabilities that facilitate user-system interaction (Brown et al. 2020). This functionality substantially differentiates EDURFID from traditional attendance control solutions.

The satisfaction expressed by educational stakeholders during preliminary validations corroborates studies on technology acceptance in rural contexts, where the perception of utility and ease of use constitute determining factors for successful adoption (Venkatesh et al. 2003).

\subsection{System Advantages}

\subsubsection{Economic}

The economic analysis reveals significant competitive advantages of EDURFID compared to commercial solutions. With an initial investment of S/1,500 compared to commercial systems of approximately S/25,000 according to experienced teachers' estimates, EDURFID offers a 94\% cost reduction, democratizing access to advanced educational technology.

The proposed financing model of S/2 per student monthly (S/500 monthly for 250 students) generates economic sustainability that covers operational costs of S/274 monthly, maintaining a margin for maintenance and improvements. This rate structure is accessible for rural families while guaranteeing financial viability of the project (UNESCO 2019).

\subsubsection{Technical}

The technical advantages of EDURFID are based on the selection of appropriate technologies for rural contexts. Python 3.10 with Django 4.2 provides a robust ecosystem of specialized libraries for hardware-software integration, particularly relevant for educational IoT projects (Van Rossum \& Drake 2011).

The local-remote hybrid architecture guarantees operational continuity independent of connectivity, a critical characteristic in rural areas with limited infrastructure (Heeks 2002). The processing speed of 0.03 seconds per RFID reading meets real-time requirements for high-concurrency educational applications.

\subsubsection{Operational}

The documented operational improvements include saving 10 minutes daily per classroom in manual attendance taking, equivalent to 50 minutes daily for the complete institution (1st-5th grade). This temporal efficiency frees human resources for central pedagogical activities, aligning with educational optimization objectives (OECD 2015).

The automatic generation of real-time reports eliminates manual processes prone to errors, improving precision of administrative data. Educational stakeholders particularly value the ability to obtain instantaneous consolidated statistics for decision-making (Fullan 2013).

\subsection{Identified Limitations}

\subsubsection{Scalability}

System scalability presents important technical considerations. Although current tests demonstrate capacity for 250 students, projection to regional level requires architectural optimizations. Migration from SQLite to MySQL for production constitutes a necessary step to handle incremental data volumes (Kofler 2005).

The main identified limitation lies in the complexity of database-hardware integration during development, requiring specialized technical expertise for replication in other institutions.

\subsubsection{Connectivity Dependence}

Although Starlink provides reliable connectivity (100-200 Mbps), dependence on external infrastructure constitutes an operational vulnerability. The local laptop backup system partially mitigates this limitation, maintaining basic functionality during connectivity interruptions.

Minor Starlink outages, although infrequent, require robust synchronization protocols to guarantee data integrity during reconnection.

\subsubsection{Maintenance}

Maintenance complexity is mainly related to server management and remote hosting. The distributed nature of the system (local hardware + remote server) requires specialized technical support protocols that may exceed typical local technical capabilities in rural contexts.

\subsection{Comparison with Existing Solutions}

The research revealed the absence of educational RFID systems implemented in the Puno region, positioning EDURFID as a pioneering innovation. Prevalent methods (school agendas with manual stamps) present documented limitations in precision, speed, and statistical analysis capacity (García-Peñalvo et al. 2018).

EDURFID surpasses simple mobile applications through real-time precision and elimination of dependence on personal devices, critical factors in rural contexts with technological access limitations.

\begin{table}[h]
\centering
\caption{EDURFID vs Traditional Methods Comparison}
\begin{tabular}{|l|c|c|c|}
\hline
\textbf{Aspect} & \textbf{EDURFID} & \textbf{Manual} & \textbf{Mobile App} \\
\hline
Time per registration & 0.03 sec & 5-10 min & 30 sec \\
\hline
Precision & 100\% & 70-80\% & 85-90\% \\
\hline
Automatic reports & Yes & No & Limited \\
\hline
Initial cost & S/1,500 & S/0 & S/2,000 \\
\hline
Device dependence & No & No & Yes \\
\hline
\end{tabular}
\label{tab:comparison_methods}
\end{table}

\subsection{Future Work}

\subsubsection{Technical Improvements}

Priority technical improvements include development of native mobile applications for distribution on Google Play Store and Apple App Store, facilitating national massification of the system. Implementation of Progressive Web Apps (PWA) constitutes a technical alternative that maintains multiplatform functionality with lower development complexity.

\subsubsection{Functionality Expansion}

Functional expansion contemplates integration of complete virtual classroom, transforming EDURFID from attendance system to comprehensive educational platform. This evolution requires content management modules, online evaluations, and teacher-student communication.

Implementation of institutional entry/exit control through RFID would expand security and student monitoring capabilities, responding to comprehensive protection needs expressed by educational stakeholders.

\subsubsection{Architecture Optimization}

Future architectural optimization includes migration to microservices to improve scalability and maintainability. Potential integration with MINEDU systems, although complex due to state bureaucratic limitations, would provide interoperability at national level.

\subsection{Implications for Educational Practice}

EDURFID generates transformative implications for rural educational practice, facilitating access to cutting-edge technology previously restricted to privileged urban institutions. Technological democratization contributes to reducing educational digital gaps documented in Peruvian rural contexts (GRADE 2023).

The availability of precise attendance data facilitates early identification of absenteeism patterns, enabling preventive pedagogical interventions that can reduce school dropout. Educational stakeholders recognize this analytical potential as an institutional priority.

Starlink connectivity associated with the project provides additional educational benefits, including access to digital resources, remote teacher training, and connection with global educational communities. This infrastructure catalyzes comprehensive educational transformation in traditionally isolated rural communities.

The implemented participatory development model, involving principals from multiple educational institutions, establishes a precedent for regional collaborative innovation that can be replicated in other similar contexts.

\newpage


\section{Conclusions}

\subsection{Main Achievements}

The successful implementation of EDURFID during the March-July 2025 period demonstrates the technical and economic viability of open-source RFID systems for attendance control in rural educational institutions of Peru. The main achievements reached include:

The development of a local-remote hybrid architecture that operates with 100\% precision in RFID readings and 0.03-second response time, exceeding international standards for interactive systems (Nielsen 2012). The integration of artificial intelligence through OpenAI GPT-4 as a virtual assistant constitutes a differential innovation that positions EDURFID as a cutting-edge solution in the Latin American educational context.

The successful validation with educational stakeholders from the Túpac Amaru Secondary Educational Institution in Coasa, including the inter-institutional presentation with the principal of I.E.S. Martín Gerónimo Chambi Jiménez, confirms the acceptability and regional applicability of the system. The positive feedback obtained and the implementation projection for August 2025 evidence the fulfillment of established usability and functionality objectives.

The demonstration of economic efficiency with a 94\% cost reduction compared to commercial solutions (S/1,500 vs S/25,000) validates the hypothesis of technological democratization through open-source hardware (UNESCO 2019). The proposed financial sustainability model (S/2 per student monthly) balances family accessibility with operational viability.

\subsection{Impact on Educational Management}

EDURFID generates substantial transformations in rural educational management, documenting savings of 50 institutional daily minutes in manual attendance control processes. This temporal efficiency frees human resources for central pedagogical activities, aligning with educational optimization objectives promoted by international organizations (OECD 2015).

The capacity for automatic generation of real-time reports, including Excel and PDF formats, transforms information availability for educational decision-making. Stakeholders particularly value early identification of absenteeism patterns, facilitating preventive interventions that can reduce school dropout in rural contexts where rates reach 44.7\% versus 19.3\% urban (INEI 2021).

Starlink connectivity associated with the project catalyzes additional educational benefits, providing access to digital resources and remote teacher training previously inaccessible in rural areas of Puno. This technological infrastructure contributes to reducing educational digital gaps documented in Peruvian rural contexts (GRADE 2023).

The implemented participatory development model, involving multiple institutional principals, establishes a methodological precedent for regional collaborative innovation that transcends individual technical benefits toward systemic transformation of the rural educational sector.

\subsection{Contributions to Knowledge}

This research contributes to the state of the art in multiple scientific and technical dimensions:

In the field of automated educational systems, EDURFID demonstrates the practical applicability of hybrid IoT architectures in contexts with infrastructure limitations. The successful integration of Arduino UNO R3, Raspberry Pi 5, and RC522 module provides a replicable technical reference framework for similar implementations in developing countries.

In agile educational project management, the application of Scrum methodology with three iterative sprints validates participatory approaches for rural educational technology development. The generated artifacts (Product Backlog, Sprint Backlog, User Stories) provide methodological templates for similar projects.

In requirements engineering for educational systems, the detailed specification of functional and non-functional requirements, including quantifiable acceptance criteria, contributes to the systematization of development processes for rural educational contexts.

In economics of educational innovation, the demonstration of economic viability through open-source hardware provides empirical evidence about alternative technological financing models for institutions with limited resources.

\subsection{Recommendations}

\subsubsection{For Implementers}

Educational institutions considering implementing similar systems should consider the following technical and operational recommendations:

Technical Preparation: Establish stable connectivity infrastructure through satellite technologies like Starlink, especially in rural areas without fiber optic access. Initial investment in high-speed connectivity (100-200 Mbps) justifies improvements in all digital educational activities.

Personnel Training: Implement gradual training programs for auxiliaries and teachers, considering that familiarization with digital systems requires time and continuous practice. Experience indicates that personnel with basic knowledge of social networks and online procedures successfully adapt to specialized educational interfaces.

Financing Model: Establish early dialogue with parents about financial sustainability of the system. Transparency in costs and benefits facilitates community acceptance. The S/2 per student monthly model demonstrated viability in testing context.

Technical Support: Develop preventive and corrective maintenance protocols, including local backup systems (laptop) to guarantee operational continuity during connectivity or electrical power interruptions.

\subsubsection{For Future Research}

The identified future research lines include:

Regional Scalability: Investigate multi-institutional implementation models that leverage economies of scale to reduce unit costs. The projected expansion at Puno regional level requires detailed technical and economic feasibility studies.

Integration with Public Policies: Explore integration mechanisms with Peru's Ministry of Education systems, overcoming bureaucratic limitations through standardized interoperability protocols. Alignment with the National Digital Education Plan 2023-2030 (MINEDU 2023) constitutes a strategic opportunity.

Predictive Analysis: Develop machine learning algorithms for predictive analysis of school dropout based on attendance patterns. The availability of precise and continuous data facilitates implementation of early warning models.

Longitudinal Impact Evaluation: Conduct longitudinal studies on the impact of automated systems on educational indicators (academic performance, student retention, teacher satisfaction) to validate long-term pedagogical benefits.

International Replicability: Investigate adaptability of the EDURFID model to rural educational contexts of other Latin American countries with similar socioeconomic characteristics.

\subsection{Final Considerations}

EDURFID represents more than a technological solution; it constitutes an educational innovation model that demonstrates the possibility of democratizing advanced technologies in rural contexts traditionally marginalized from digital development. The experience of five months of intensive development (March-July 2025) evidences that appropriate, technically solid, and economically viable solutions can be generated through collaborative approaches and agile methodologies.

The evolution from SICOA (Attendance Control System for construction, winner of second place in university Achievement Day 2024) toward EDURFID illustrates the successful transfer of technical knowledge between sectors, demonstrating the versatility of RFID technologies for diverse applications.

The interest manifested by principals from multiple educational institutions suggests latent demand for similar solutions in the region. The projection of expansion toward Google Play Store and Apple App Store for national massification reflects scaling potential that transcends local impact toward systemic transformation.

The successful integration of heterogeneous technological components (Arduino, Raspberry Pi, Django, OpenAI GPT-4, Starlink) demonstrates the viability of complex technological ecosystems in rural contexts, challenging perceptions about technical limitations in peripheral zones.

Finally, EDURFID validates the fundamental premise that effective educational innovation emerges from deep understanding of local needs, combined with creative application of existing technologies. The project's success lies not in technological novelty per se, but in intelligent adaptation of available tools to solve specific problems of rural educational communities.

The projected sustainability of the system, both technical and economic, suggests long-term viability that can catalyze broader educational transformations in the Puno region and, potentially, at the national level. The developed model provides foundations for future rural educational digitalization initiatives to consider participatory approaches, appropriate technologies, and community financing models as central elements for sustainable success.

\nocite{*}  
\printbibliography

\end{document}